 \documentclass[journal]{IEEEtran}

 \usepackage[left=0.625in,right=0.625in,top=0.75in,bottom=0.99in]{geometry}
\usepackage{cite}
\usepackage{amsmath,amssymb,amsfonts}
\usepackage{graphicx}
\usepackage{color}
\usepackage{subfig}
\usepackage{stackengine}
\usepackage{comment}
\usepackage[normalem]{ulem}
\usepackage{tabularx}
\usepackage{bm}
\usepackage{dsfont}
\usepackage{mathtools}
\usepackage[font=scriptsize]{caption}

\usepackage{booktabs,multirow,array,caption}

\usepackage[singlespacing]{setspace} 
\usepackage[colorlinks, citecolor=blue, bookmarks=false]{hyperref} % to highlight references with colors

\usepackage[flushleft]{threeparttable} % http://ctan.org/pkg/threeparttable
\usepackage{booktabs}

\usepackage{textcomp}
\usepackage{xcolor}

\usepackage[algo2e]{algorithm2e} 
\SetKwRepeat{Do}{do}{while}%

\SetCommentSty{mycommfont}
\usepackage{algorithm}
\usepackage{algpseudocode}

\def\BibTeX{{\rm B\kern-.05em{\sc i\kern-.025em b}\kern-.08em
    T\kern-.1667em\lower.7ex\hbox{E}\kern-.125emX}}
    
\algnewcommand\algorithmicforeach{\textbf{for each}}
\algdef{S}[FOR]{ForEach}[1]{\algorithmicforeach\ #1\ \algorithmicdo}

\usepackage{orcidlink}

\usepackage{xspace}
\usepackage{ifthen}
\usepackage[inline]{enumitem}

\usepackage[utf8]{inputenc}

\usepackage{algcompatible}
\usepackage{algorithmicx}

\newboolean{showcomments}
\setboolean{showcomments}{true}
\ifthenelse{\boolean{showcomments}}
{
}

\usepackage{subfig}
\usepackage{fancyhdr}
\usepackage{amssymb}% http://ctan.org/pkg/amssymb
\usepackage{pifont}% http://ctan.org/pkg/pifont

\newcommand{\cmark}{\textcolor{green}{\ding{51}}} % Green check mark
\newcommand{\xmark}{\textcolor{red}{\ding{55}}}   % Red cross mark

\begin{document}

\renewcommand{\algorithmiccomment}[1]{\textit{#1}}% Updated definition

%\title{\huge LLM-Driven Deployment of 6G-Enabled Internet of Automated Defense Vehicles: Opportunities and Challenges}
\title{\huge Leveraging Edge Intelligence and LLMs to Advance 6G-Enabled Internet of Automated Defense Vehicles}

\author{
\IEEEauthorblockN{Murat Arda Onsu, Poonam Lohan, Burak Kantarci}

\thanks{The authors are with the School of Electrical Engineering and Computer Science University of Ottawa,
Ottawa, Canada 
\{monsu022, ppoonam, burak.kantarci\}@uottawa.ca}

}

\maketitle

\begin{abstract}
The evolution of Artificial Intelligence (AI) and its subset Deep Learning (DL), has profoundly impacted numerous domains, including autonomous driving. The integration of autonomous driving in military settings reduces human casualties and enables precise and safe execution of missions in hazardous environments while allowing for reliable logistics support without the risks associated with fatigue-related errors. However, relying on autonomous driving solely requires an advanced decision-making model that is adaptable and optimum in any situation. Considering the presence of numerous interconnected autonomous vehicles in mission-critical scenarios, Ultra-Reliable Low Latency Communication (URLLC) is vital for ensuring seamless coordination, real-time data exchange, and instantaneous response to dynamic driving environments. The advent of 6G strengthens the Internet of Automated Defense Vehicles (IoADV) concept within the realm of Internet of Military Defense Things (IoMDT) by enabling robust connectivity, crucial for real-time data exchange, advanced navigation, and enhanced safety features through IoADV interactions. On the other hand, a critical advancement in this space is using pre-trained Generative Large Language Models (LLMs) for decision-making and communication optimization for autonomous driving.
Hence, this work presents opportunities and challenges with a vision of realizing the full potential of these technologies in critical defense applications, especially through the advancement of IoADV and its role in enhancing autonomous military operations. 
\end{abstract}

\begin{IEEEkeywords}Autonomous driving, IoADV, 6G-enabled Edge intelligence, decision-making, Multimodal LLMs.

\end{IEEEkeywords}

\section{Introduction} \label{sec:1}
The potential of military-grade AI is vast and multifaceted, driven by its ability to analyze large volumes of data from diverse sources. This analytical capability can be applied to various defense technologies, including drones, military vehicles, armored tanks, and weapon systems \cite{15}.

Automated defense Vehicles (ADVs) in military settings offer strategic benefits, including reducing risk to soldiers, enhancing reconnaissance, and improving logistics and supply chain efficiency in combat zones. These defense vehicles must navigate and operate in hazardous, complex environments with minimal human intervention. However, challenges arise in military settings, such as terrain awareness, off-road navigation, unfamiliar environments, the need for complete re-routing in open areas, identifying alternate routes, and ensuring optimal vehicle control \cite{14}.  To address these challenges, integrating multiple connected vehicles into the 6G-enabled Internet of Automated Defense Vehicles (IoADV) framework can significantly improve situational awareness, coordination, and operational efficiency. The 6G-enabled IoADV allows real-time data sharing, autonomous convoy operations, and synchronized maneuvers, ensuring safe and efficient missions in hostile environments. Additionally, military-grade AI faces ethical and legal challenges, including transparency in decision-making and liability for errors or collateral damage.  Thus managing large volume data, ensuring real-time processing, maintaining reliable communication, efficient decision-making, and addressing cybersecurity threats are critical issues \cite{15}.

Specifically, there are two main challenges: reliable communication and decision-making for autonomous driving of ADVs. To enhance communication, the use of a large generative AI (GenAI) model is proposed for 6G-enabled IoADV. This model leverages advanced data processing, multi-modal integration, and predictive analytics to improve communication and coordination among vehicles \cite{generative}. This approach aims to enable more reliable, efficient IoADVs by utilizing generative and predictive capabilities of AI models to optimize various network functions and support real-time decision-making. For decision-making in autonomous driving, a framework that integrates pre-trained generative Multimodal Large Language Models (M-LLMs) \cite{13} is proposed due to their advanced features to assist ADVs by integrating and processing diverse data types such as visual inputs, sensor readings, and textual information. This integration enhances vehicle's ability to understand and respond to complex driving environments, improving perception, decision-making, and real-time navigation. Moreover, LLMs provide transparency and explainability in decision-making, addressing the ``black box" issue associated with deep neural networks and other advanced AI models. To the best of our knowledge, \textit{this is the first time in the literature that LLMs are being explored within the context of 6G-enabled IoADVs for military applications.}

\section{Use Cases and Requirements for Autonomous Defense Vehicles} \label{sec:2}

\subsection{Use Cases in Military Defense Services}

ADVs are integral to modern military operations, offering a range of applications that enhance operational efficiency and safety. In a critical extraction mission, LLMs can direct autonomous vehicles to rapidly extract personnel from the field. Imagine a team of special forces surrounded by enemy combatants. ADVs equipped with advanced IoADV sensors and cameras conduct real-time intelligence gathering, providing comprehensive situational awareness without exposing troops to danger. If an ambush is detected on the primary route, the LLM quickly recalibrates the vehicle’s path, opting for a safer, albeit less direct, alternative route.

Additionally, ADVs enhance reconnaissance, surveillance, and intelligence gathering while excelling in search and rescue missions by swiftly locating individuals in distress. They bolster border and base security with constant vigilance and rapid response. User-friendly control systems enable efficient management, making them vital for modern military operations. In logistics, ADVs autonomously transport supplies and equipment through challenging terrains, ensuring timely delivery to frontline units while minimizing risks to personnel. 

Moreover, in combat support roles, armed ADVs operate alongside human troops, delivering suppressive fire and engaging enemy targets, thereby amplifying combat effectiveness and reducing personnel exposure to threats. Also, ADVs perform engineering tasks such as mine detection and clearance, forging safe pathways through hazardous areas, and safeguarding human engineers from potential explosions. In medical emergencies, ADVs swiftly evacuate wounded soldiers from the front lines, navigating under fire to deliver casualties to medical facilities, thus enhancing survival rates.

\subsection{Requirement of Military Defense Services}

%Military defense requirements encompass a wide range of technological and ethical considerations to ensure effective and secure operations. Though autonomous vehicle research is common in the civilian domain, it can not be directly integrated into the military domain. One of the reasons is civilian autonomous vehicles often lack the robust security measures required for military operations, making them susceptible to cyberattacks and electronic warfare. Adversaries can exploit unencrypted communication links to intercept data or disrupt operations through jamming and spoofing attacks. The reliance on commercial off-the-shelf technologies without adequate hardening can compromise mission integrity.

%Moreover, civilian technologies are typically designed for controlled environments and may not withstand the harsh conditions of military operations, such as extreme temperatures, dust, and electromagnetic interference. Also, integrating civilian systems with existing military infrastructure poses significant interoperability issues. Differences in communication protocols, data formats, and operational standards can hinder seamless integration, leading to inefficiencies and potential mission failures. Ensuring civilian technologies can operate cohesively with military systems requires extensive modifications and rigorous testing \cite{8682048}.

Military defense operations demand robust technological and ethical considerations, making direct integration of civilian autonomous vehicles unfeasible. These vehicles often lack the necessary security measures, leaving them vulnerable to cyberattacks, jamming, and spoofing. Additionally, their reliance on commercial off-the-shelf components compromises mission integrity due to inadequate hardening. Beyond security risks, civilian technologies are designed for controlled environments and may fail under extreme military conditions, such as harsh weather and electromagnetic interference. Interoperability challenges further complicate integration with military infrastructure, as differences in communication protocols and operational standards require extensive modifications and rigorous testing \cite{8682048}.

\begin{figure} [t]
    \centering
    \includegraphics[width=\linewidth]{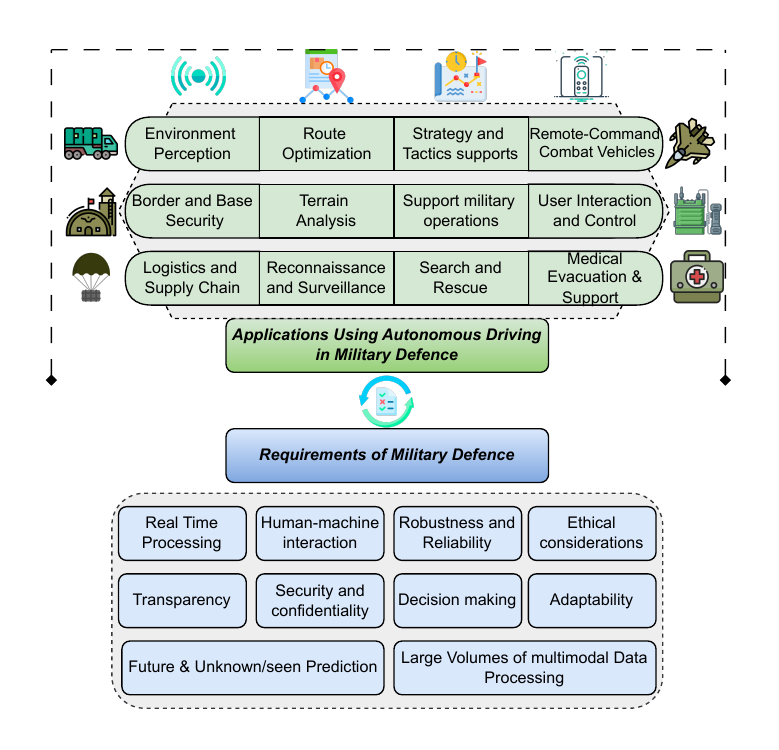}
    \caption{Military Applications and Requirements for Autonomous Defense Vehicles}
    \label{fig:img2}
\end{figure}

On the other hand, real-time processing is critical for immediate response and decision-making in dynamic combat scenarios. Since it is possible to operate in an unfamiliar environment, the ability to predict future and unknown events enhances strategic planning and preparedness. Also, robustness and reliability are essential to maintain operational continuity in harsh and unpredictable environments. Effective human-machine interaction ensures seamless integration of autonomous systems with human operators, enhancing overall mission efficiency. Compared to civilian technologies, mistakes and faults have much worse consequences for the military domain, such as the risk of losing supply, life, or operation, so all these challenges should be considered before involving autonomous vehicles in action. Overall requirements and potential military defense applications of autonomous driving are illustrated in \figurename \ref{fig:img2}. 

To integrate autonomous driving into the military domain, AI methodologies enhance data analysis, decision-making, and automation. Military-grade AI focuses on defense-specific technologies, with recent reliance on DL methods to improve decision-making \cite{14}. However, DL models face challenges such as a lack of transparency, making validation and traceability difficult, and requiring extensive training data, which is hard to obtain due to dynamic tactics and environments. Additionally, optimizing network performance is crucial for reliable communication and resilience to adversarial attacks, particularly in multi-vehicle, real-time coordination scenarios. 

\section{6G-enabled Edge intelligence and IoADV} \label{sec:3} % 6G V2X

Communication systems and network applications are vital in military operations, underpinning mission-critical areas such as time-sensitive targeting, covert special operations, command and control, training, and logistics. These domains rely on high-capacity access, seamless mobility, and robust transmission to function effectively. The integration of advanced communication technologies such as 6G and Edge Intelligence ensures that military vehicles can coordinate efficiently, maintain situational awareness, and execute operations with precision across diverse and dynamic environments.

\subsection{Enhanced Communication in IoADV via 6G and Edge Computing}

\begin{figure} [t]
    \centering
    \includegraphics[width=0.95\linewidth]{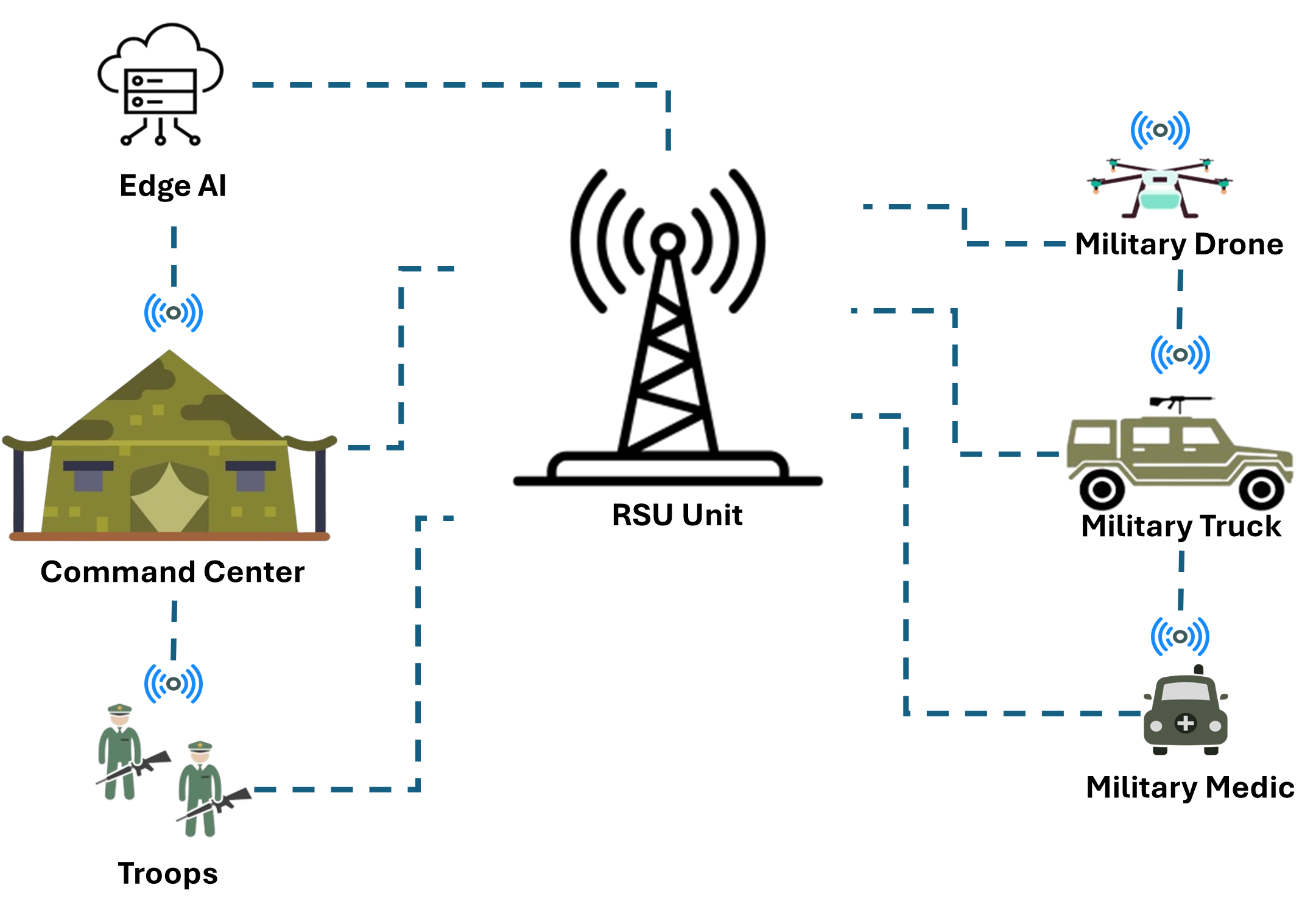}
    \caption{IoADV in Military Domain}
    \label{fig:img3}
\end{figure}

IoADV technology, a sophisticated network of interconnected vehicles and systems within the military domain, enhances combat operations by improving real-time data sharing and situational awareness. Integrating IoADV into military ground vehicles enables continuous health monitoring through sensors, supporting predictive maintenance, and minimizing downtime. It also provides ground forces with timely updates on vehicle conditions and environmental factors for informed decision-making. Moreover, IoADV enhances advanced navigation and communication systems, ensuring seamless unit coordination and boosting mission efficiency. This network integrates sensors, unmanned vehicles, weapons systems, and wearable technologies, enabling effective environmental interaction and coordination of activities. The proposed communication model is depicted in \figurename \ref{fig:img3}. 

The 5G model falls short in meeting the connectivity demands of autonomous vehicles, especially in terms of seamless reliable communication. In contrast, 6G-based IoADV enhances safety and supports user needs with advanced technologies such as THz communication, Non-orthogonal Multiple Access (NOMA), Cell-free Massive MIMO, MEC, AI, and ML, enabling ultra-low latency ($0.1$ ms) and $99.99999\%$ reliability for real-time decision-making. With data rates up to 1 Tbps, 6G facilitates rapid transmission of high-definition maps and sensor data \cite{7}. AI and edge computing further improve autonomous vehicle intelligence by processing data at the edge, minimizing latency and improving response times  \cite{friha.comst.2024}. The integration of 6G and edge computing reduces latency significantly compared to cloud-based 5G networks, which experience total latencies between $20$ and $70$ ms. Incorporating edge computing into a 5G or hybrid 6G network reduces latency to $6$–$15$ ms, with the greatest improvement seen in 6G and edge computing integration, achieving latencies under $5$ ms. This includes processing latencies below 1 ms and network latencies between $1$–$5$ ms, making it ideal for real-time, efficient decision-making, especially in military applications.

\begin{figure} [t]
    \centering
    \includegraphics[width=0.95\linewidth]{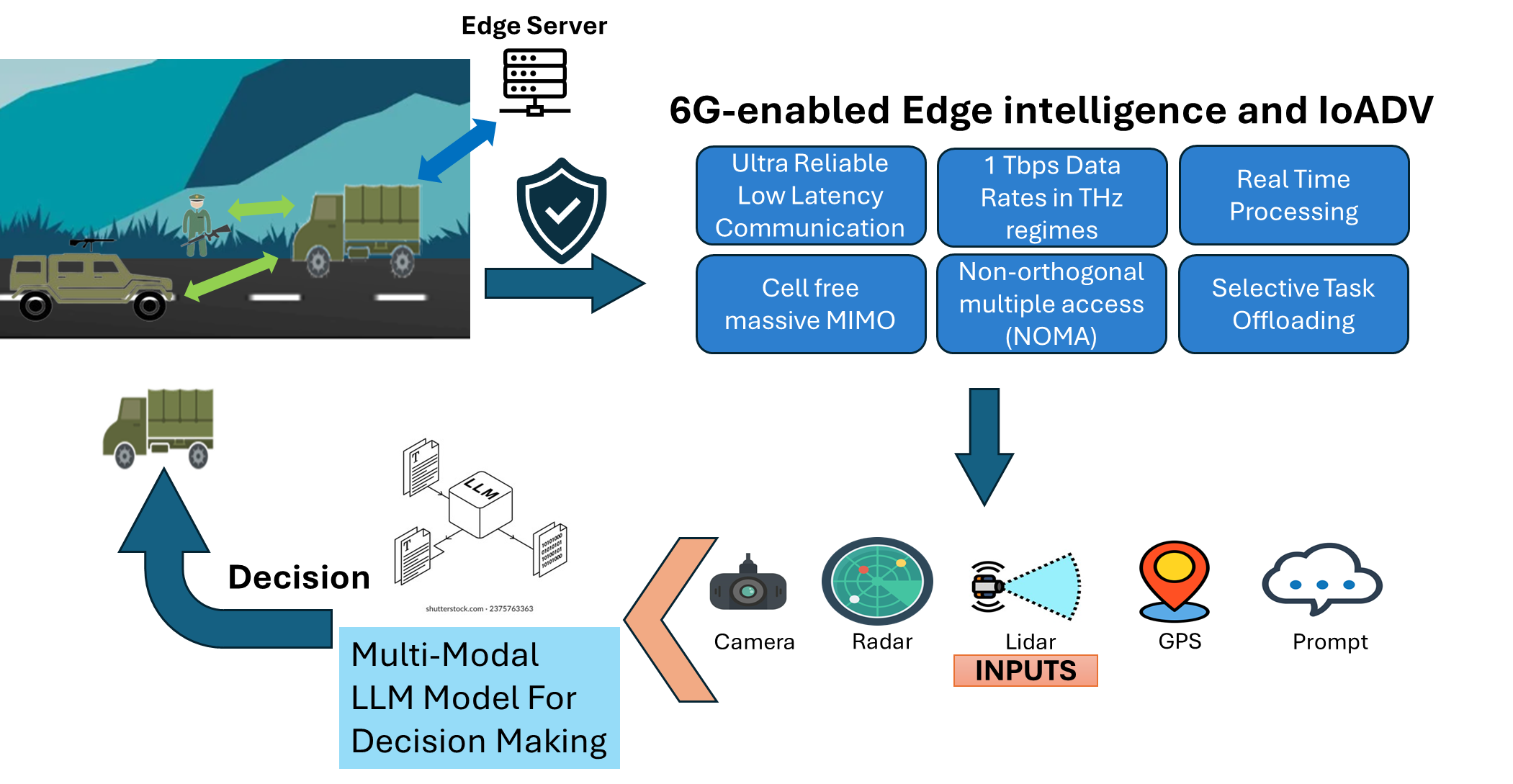}
    \caption{Proposed Solution Workflow for 6G-enabled Edge intelligence and IoADV}
    \label{fig:img4}
\end{figure}

\subsection{Security Concern of Communication}

The increased connectivity and multi-layered architecture of IoT expose systems to cybersecurity risks, where a breach in one device can compromise the entire network. In military contexts, compromised IoADV devices risk unauthorized access to sensitive data, manipulation, and communication disruption. These devices face unique challenges due to their critical nature and hostile environments \cite{pasdar2024cybersecurity}. Robust security measures are crucial for protecting sensitive data and maintaining operational integrity. Military communication systems can enhance resilience using distributed learning models and cryptographic techniques. Semantic edge-based systems improve scalability, while lightweight access control protocols ensure secure communication via multi-factor authentication. Threat modeling (e.g., STRIDE, DREAD) helps identify vulnerabilities, and decision control frameworks, like emotion-aware systems and multi-level data fusion, enhance battlefield awareness and real-time decision-making \cite{pasdar2024cybersecurity}.%Strong security measures are essential to protect sensitive information and maintain operational integrity. There are several suggestions to overcome these security challenges in IoT communication throughout the literature. To enhance trust and resilience, military communication systems can adopt distributed learning models and cryptographic techniques, ensuring adaptability and reliability. Advanced network protocols like semantic edge-based systems improve scalability and interoperability, while lightweight access control protocols secure communication through multi-factor authentication and session key establishment. Threat modeling techniques, including STRIDE and DREAD, enable systematic identification and mitigation of vulnerabilities. Finally, decision control frameworks, such as emotion-aware systems and multi-level data fusion, enhance battlefield situational awareness and support real-time, precise operational decisions \cite{pasdar2024cybersecurity}.

\section{LLM Integration for Communication and Decision Making} \label{sec:4}
Language modeling (LM) is an essential method for improving machine language intelligence by modeling the generative likelihood of word sequences to predict future tokens. LLMs, like the 175-billion parameters GPT-3, are scaled versions of Pre-Trained Language Models (PLMs). These models can also be fine-tuned for specific tasks with minimal data. The backbone of LLMs typically relies on transformer architecture, which uses self-attention mechanism to weigh important contexts, handle long-range dependencies, and understand nuanced meanings. The primary method of accessing LLMs is through prompting interface, which provides insight into their operation, enhances explainability, and formats their tasks in a way that can be easily followed \cite{4}.

\subsection{Network Optimization Using LLM and Generative AI} \label{sec:4.1}

LLMs and GenAI significantly impact the field of telecommunications by enhancing wireless communication and enabling more intelligent, autonomous networks. These models improve situational and contextual awareness by leveraging multimodal data, such as radio frequency (RF) signals and 2D/3D representations of wireless environments. Their predictive capabilities allow networks to proactively manage tasks like localization, beamforming, power allocation, and spectrum management. Another benefit of GenAI is to excel in cyber threats such as prompt injection, insecure output handling, data poisoning, DDoS attacks, jamming, and so on.

A significant advancement in GenAI is its ability to enable a vast array of wireless IoADV devices to provide collective intelligence. In 6G, computing infrastructure that facilitates the seamless transfer of knowledge across the network should be created. To accomplish this, large GenAI models must be anchored in real-world contexts and able to communicate knowledge to perform multi-agent planning, decision-making, and reasoning. For this part, key research perspectives can be classified into three categories: Semantic communication, Emergent protocol learning, and Distributed Large-GenAI-Models powered AI agent \cite{generative}.

Large GenAI models generate massive raw data, leading to high computational overhead and redundancy. Semantic communication mitigates this by transmitting only the intent and meaning behind data, reducing transmission size and optimizing network throughput. This requires redesigning the wireless physical layer to enable knowledge-based learning from compressed, abstracted data. By prioritizing mission-critical information, such as enemy positions in military operations, semantic communication enhances resource utilization and secures transmissions. Moreover, Emergent protocol learning allows AI agents to autonomously develop adaptive communication protocols using Multi-Agent Reinforcement Learning (MARL), enabling context-aware MAC and network optimization. Large GenAI models enhance MARL by guiding collaborative actions, reducing data needs, and improving convergence. Distributed GenAI models further streamline real-world task planning by coordinating actions and delegating sub-tasks, improving efficiency through cooperative intelligence.

%For instance, during coordinated military tank strike, Multi-agent reinforcement learning (MARL)-trained agents can dynamically adjust protocols to optimize resource allocation for navigation, target acquisition, and secure data exchange. Large GenAI models act as guides for MARL, offering expert planning capabilities that improve convergence and reduce the need for extensive data collection. Another example of these capabilities in action is distributed military reconnaissance mission involving drones and ground units. In such scenario, distributed large-GenAI models can create action sequences for the collaborative vehicles, assigning tasks like mapping enemy territories, identifying threats, and transmitting findings to command units. By leveraging real-world grounding and cooperative planning, they can operate autonomously, sharing critical insights while minimizing redundancy. This approach enhances mission execution efficiency, improves resource utilization, and ensures seamless coordination between all units involved in the operation. \cite{generative_cyber}

During a coordinated military tank strike, MARL-trained agents dynamically adjust protocols for navigation, target acquisition, and secure data exchange, while large GenAI models enhance planning, improving convergence and reducing data requirements. In distributed reconnaissance missions, GenAI models coordinate drones and ground units, assigning tasks like mapping enemy territories, identifying threats, and transmitting intelligence. By leveraging cooperative planning and real-world grounding, they enable autonomous operations, optimizing resource use and ensuring seamless coordination \cite{generative_cyber}.

On the other hand, LLMs are crucial in countering cybersecurity threats by leveraging natural language understanding, pattern recognition, and decision-making abilities. They excel at real-time threat detection and analysis, particularly against jamming signals that pose significant risks to ADVs, which depend on wireless communication for receiving commands and applying decision-making. While Frequency Hopping Spread Spectrum (FHSS) is a common anti-jamming technique, military environments require more adaptive solutions due to the higher stakes of jamming attacks. LLMs can enhance the security of command-and-control systems by monitoring and detecting advanced threats like Advanced Persistent Threats (APTs), phishing campaigns, jamming, and DDoS attacks, and automating responses to protect critical resources. In anti-jamming protocols, LLMs dynamically adapt by analyzing attack types and severity, optimizing frequency hopping, adjusting beamforming, and fine-tuning power levels to ensure robust and resilient communication.

%LLMs analyze vast amounts of network data collected in IoADV sensors in real-time, identifying anomalies and recognizing patterns indicative of cybersecurity threats, such as malware, jamming signals, and unusual traffic behaviors. Their ability to process and contextualize large datasets allows them to flag potential issues before they escalate, outperforming traditional models in areas like malware detection and intrusion detection \cite{generative_cyber}.

%In a military operation, command and control systems are the backbone of effective decision-making and coordination, often functioning in dynamic and high-risk environments. These systems are vulnerable to various cyber threats, including Advanced Persistent Threats (APTs), phishing campaigns, jamming, and DDoS attacks. LLMs can play a crucial role in securing these systems by providing real-time monitoring, detecting sophisticated cyber threats, and automating responses to maintain the integrity and availability of critical resources. \noteArda{Moreover, incorporating LLMs with advanced AI models into anti-jamming protocols represents another approach that can interpret complex communication patterns and environmental factors, enabling informed decision-making to mitigate jamming.}

\begin{figure} [t]
    \centering
    \includegraphics[width=1.0\linewidth]{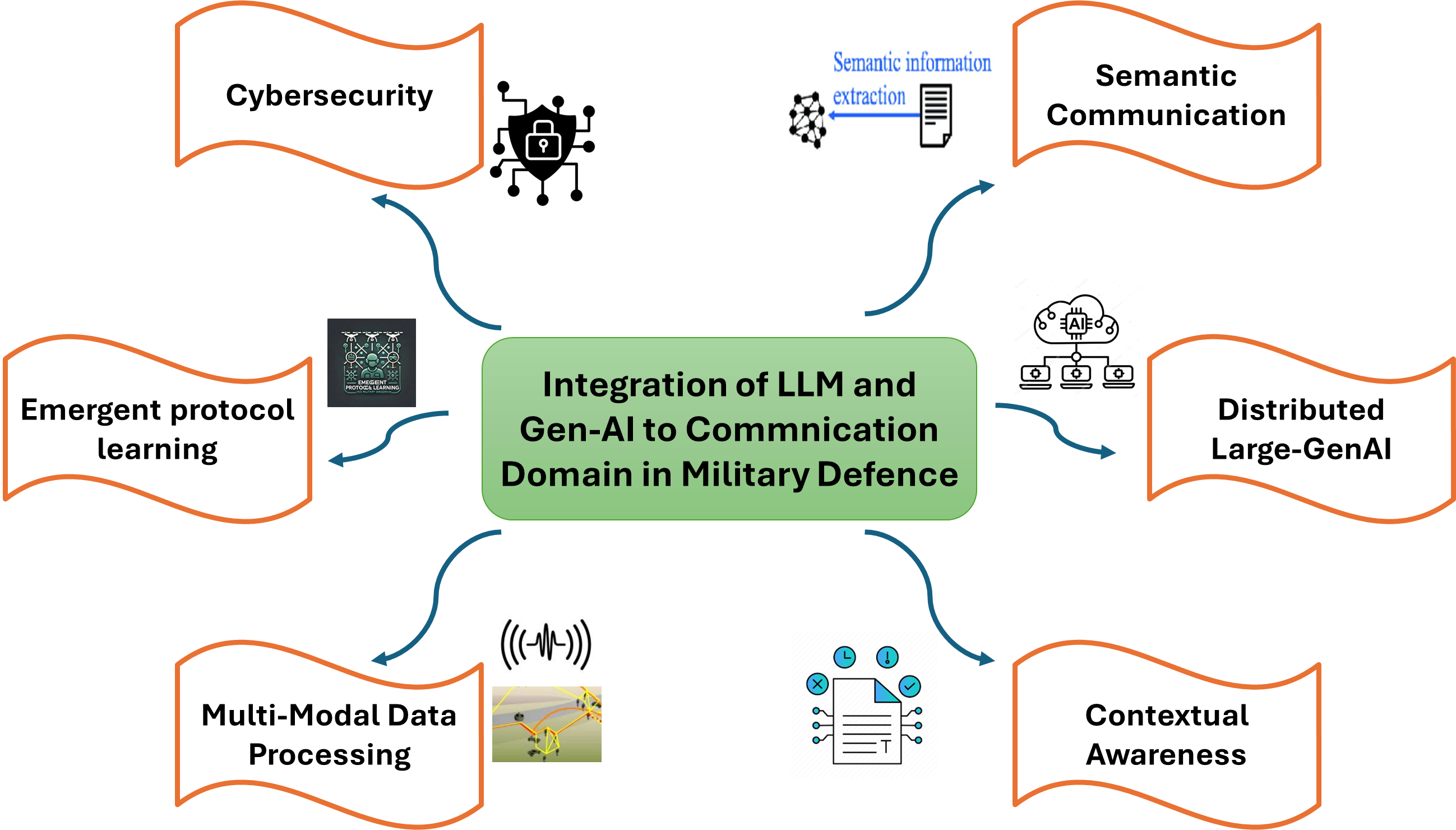}
    \caption{LLM Assistance for Communication in IoADV in Military Domain}
    \label{fig:img5}
\end{figure}

For instance, consider a military ground vehicle operation relying on secure networks for command and data transmission, an adversary attempts or cyberattacks intercept vehicle communications or redirect its mission. The LLM identifies anomalies in network traffic suggesting an unauthorized access attempt. It analyzes the threat and recommends immediate actions, such as encrypting the data channel and isolating the compromised node. Simultaneously, it sends alerts to operators and provides a detailed analysis of the attack’s origin, aiding in counterintelligence efforts. The general impacts of LLM on communication and military defense are illustrated in \figurename \ref{fig:img5}.

\subsection{Decision-Making in ADVs by Multimodal-LLM} \label{sec:4.2}

Though LLMs can provide efficient communication and security in IoADV, it is also essential to enhance individual and connected defense vehicles' actions and decisions in critical military operations. ADVs can also utilize multimodal-LLM framework for advanced decision-making and interaction with the environment. Multimodal data combines information from various IoADV sensors attached to the connected vehicles, such as cameras, LiDAR, Radar, and GPS, to form a comprehensive understanding of the vehicle's surroundings. This integration significantly enhances the perception capabilities of autonomous driving systems. Through the combination of visual data from cameras with accurate distance measurements from LiDAR and Radar, the system enhances its ability to identify and categorize objects, calculate their velocity, and forecast their paths.

Moreover, multimodal data enables redundancy and reliability in perception. Different IoADV sensors have varying strengths and weaknesses; for example, cameras provide high-resolution imagery but can be affected by lighting conditions, while Radar is less detailed but performs well in adverse weather. Combining these data sources ensures that an AV maintains situational awareness even if one sensor's performance is compromised. This redundancy enhances the vehicle's ability to handle a wide range of scenarios, from bright sunlight to heavy rain or fog, ensuring safer and more reliable operation. For instance, in military operations, an autonomous combat vehicle equipped with multimodal sensors can navigate a battlefield obscured by smoke or dust from explosions, using thermal imaging to detect hidden threats or Radar to identify enemy movements through visual obstructions. This capability is critical for mission success in dynamic and hostile environments.

\subsubsection{Advancement Over Traditional Methods}

Conventional autonomous driving systems rely on rule-based algorithms for route planning but struggle in unforeseen scenarios like rare accidents, surprise attacks, or unknown territory. Advanced DL models outperform these methods in planning and decision-making but face challenges due to their black-box nature, raising ethical and legal concerns \cite{13}. Issues such as model illusions can cause misinterpretations of the environment, leading to safety risks, while biases may result in unfair decisions across diverse environments. %Additionally, errors in reasoning and false information can lead to inappropriate or dangerous driving behaviors. Moreover, inductive advice might expose vehicles to external interference or malicious actions \cite{12}.

\begin{figure} [t]
    \centering
    \includegraphics[width=0.95\linewidth]{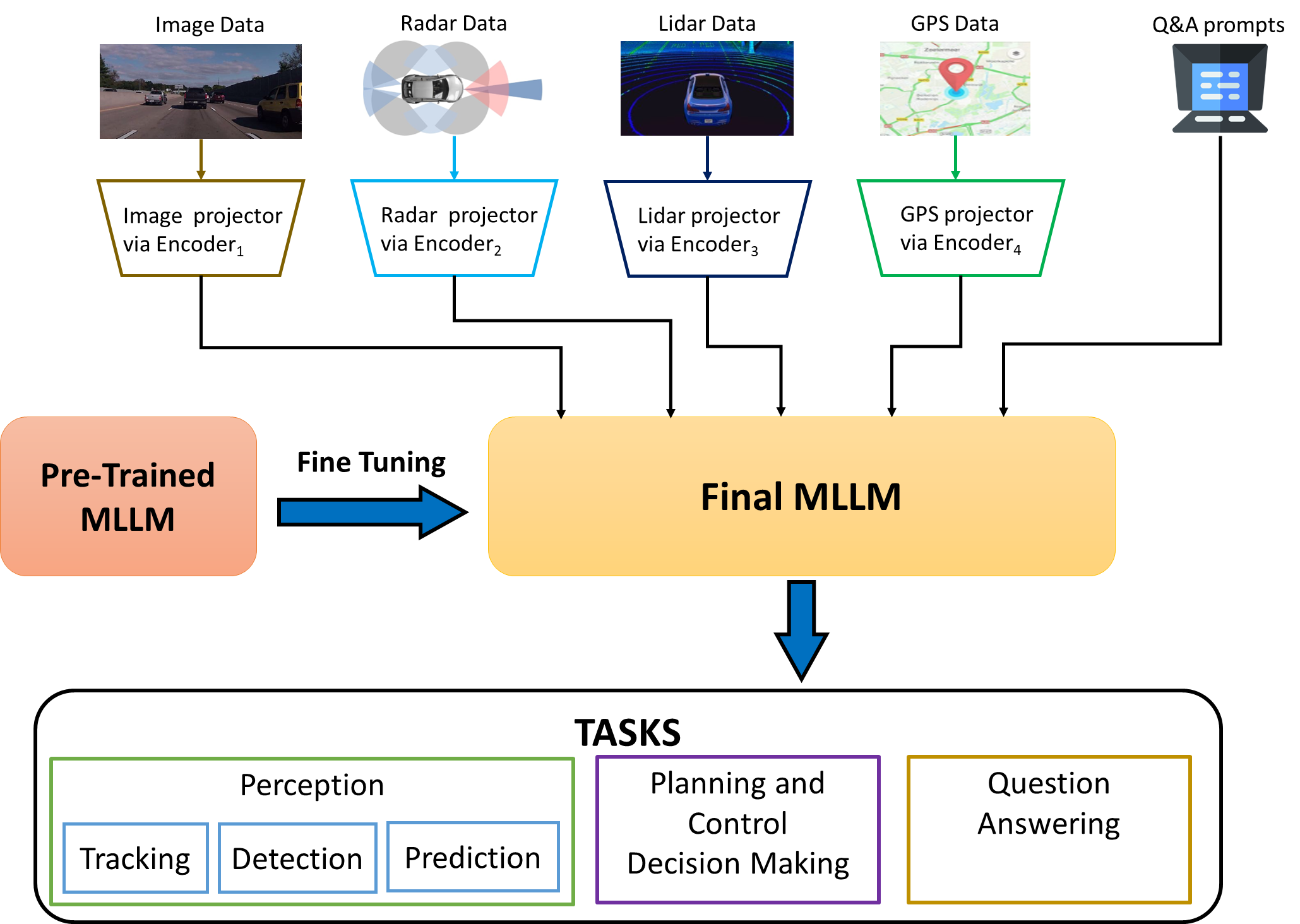}
    \caption{Multimodal-LLM Workflow for Decision Making in Automated Driving}
    \label{fig:img6}
\end{figure}

Explainable autonomous driving is crucial for clarifying the ``black box" nature of decision-making. Multimodal LLMs enhance explainability by processing human-like language, improving vehicle-person interaction, and integrating contextual data from roads, civilians, weather, and obstacles. For military applications, fine-tuned LLMs adapt to domain-specific knowledge, including doctrines, mission objectives, and foreign environments, addressing gaps in traditional training datasets. They also facilitate continuous learning by analyzing driving scenarios, enemy territories, and operational feedback. In critical decisions, such as target engagement, LLMs assess compliance with ethical standards, distinguishing combatants from non-combatants while ensuring proportionality and necessity. Additionally, LLM-assisted ADVs enhance military operations by providing real-time feedback and decision support. The proposed framework's workflow is shown in \figurename \ref{fig:img6}.

\subsubsection{Ethical Consideration of LLM}

Addressing ethical concerns in military operations necessitates adherence to established frameworks and protocols that ensure compliance with international standards. A structured framework is proposed to evaluate the ethical implications of emerging technologies in military contexts in research \cite{ethicframewrok}. This framework emphasizes the importance of aligning technological developments with established ethical principles and international standards, such as the Law of Armed Conflict (LOAC). By systematically assessing factors such as proportionality, distinction, and necessity, the framework aims to ensure that new technologies are integrated into military operations responsibly and ethically. The authors advocate for proactive ethical evaluations to anticipate potential dilemmas and guide the development and deployment of technologies in a manner consistent with both national and international ethical obligations.

\subsubsection{Literature Review of Multimodal LLM and Autonomous Vehicles}

Several studies have enhanced LLM performance in autonomous driving (Table \ref{my table}). Most use images and user prompts, while GPS and radar data were incorporated in \cite{13}. Lifelong learning in \cite{11} enables continuous adaptation, allowing the system to refine its knowledge over time. The DriveLLM framework in \cite{13} employs Reinforcement Learning (RL) for self-reflection, assessing decisions, and summarizing mistakes. Its training follows a two-tier RL process, learning from both cyber and physical environments with iterative feedback. It achieved a 0.09s median decision-making time and 5–13 decisions per second (DPS) using GPT-3.5 Turbo. Similarly, RL in \cite{11} improves continuous learning by evaluating past decisions and refining future behavior. Performance metrics include lifelong learning effectiveness and collaboration success rate, with success steps (SSmean) increasing from 8.2 to 17.6 in intersections using memory shots. Communication and reflection modules further boosted collaboration success from 50\% to 80\%, demonstrating adaptive learning benefits.

LLMs enhance resilience in adversarial military scenarios by mitigating risks and reducing latency. HUDSON \cite{n2} strengthens autonomous driving against perception attacks using multimodal inputs (images, RADAR/LIDAR) and LLM-based reasoning. It improves object detection, motion planning, and decision-making, ensuring robustness against adversarial manipulations. Evaluations under simulated attacks assess attack success rate, detection accuracy, decision correctness, and adversarial robustness. The performance metrics include attack success rate, object detection accuracy, decision correctness in motion planning, adversarial robustness, and the impact of perception errors on driving outcomes. Work \cite{NN22} addresses long-tail driving scenarios by leveraging LLMs for behavior planning, trajectory optimization, and safety verification. Experiments show LLM-driven decision-making enhances safety, computational efficiency, and driving performance. Compared to the DriveLikeAHuman baseline, it ensures non-collision behavior in 1500 steps, improves average speed (34.3±7.7 m/s vs. 25.0±5.1 m/s), and significantly reduces latency per decision (1.7±2.7s vs. 31.1±12.2s). performance metrics include safety assurance, time efficiency in terms of computational latency measured by LLM's response time, and driving performance metrics, including average speed. Long-tail scenarios, often rare and complex, challenge traditional models due to limited training data, but LLMs offer adaptive solutions. In military operations, vehicles mostly encounter this kind of situation where roads and buildings are deconstructed or terrible weather conditions, and adversarial attacks mostly happen during the decision-making in the operations. These works provide better decision-making under uncertainty, robustness for adaptation, and faster actions for ADV.

%Furthermore, research \cite{12} is performed in various datasets and simulation environments, such as CARLA, and Combines LLMs with vision-based models. Tasks include planning, perception, and reasoning with multimodal inputs such as text, images, and LiDAR. One of the datasets, such as DRAMA, stands out as a comprehensive resource with 91 hours of data, 17,785 videos, and 77,639 QA pairs, emphasizing risk localization and visual reasoning for autonomous driving. It sets a benchmark for improving decision-making and reasoning capabilities.

Moreover, a memory module is used in the framework to recall past experiences and make better decisions. The cognitive memory module stores different types of memories, such as commonsense knowledge, rules, and guidance. Therefore, this commonsense memory ensures that the agents have a foundational understanding of driving principles, which helps in making informed decisions. Experience memory and reflection memory are other types of memory that contain records of past driving scenarios, feedback, and lessons learned from previous actions. These help the AV to recall relevant memories to inform decision-making, thus enabling the agent to learn from past interactions and experiences. 

\begin{table}[t] 

\fontsize{8.0}{9.7}\selectfont

\caption{Comparison of Recent Automated Driving Research using LLM \\(\cmark: Yes/Mentioned \xmark: No/Not Mentioned)}

\begin{tabular}{llccccc}
\hline
\textbf{Work}                     & \multicolumn{1}{c}{}                                                   & \cite{1}        & \cite{NN22}     & \cite{11}       & \cite{n2}       & \cite{13}       \\ \hline
\multirow{4}{*}{\textbf{Dataset}} & Image                                                                  & \textit{\cmark} & \textit{\cmark} & \textit{\cmark} & \textit{\cmark} & \textit{\cmark} \\ \\
                                  & Radar/Lidar                                                            & \xmark          & \xmark          & \xmark          & \textit{\cmark} & \textit{\cmark} \\ \\
                                  & GPS                                                                    & \xmark          & \xmark          & \xmark          & \xmark          & \textit{\cmark} \\ \\
                                  & Prompt                                                                 & \textit{\cmark} & \textit{\cmark} & \textit{\cmark} & \textit{\cmark} & \textit{\cmark} \\ \hline
\textbf{Communications}           & 5G/6G                                                                  & \xmark          & \xmark          & \xmark          & \xmark          & \textit{\cmark} \\ \\
\textbf{}                         & MEC                                                                    & \xmark          & \xmark          & \xmark          & \xmark          & \textit{\cmark} \\ \hline
\multirow{9}{*}{\textbf{Tasks}}   & \begin{tabular}[c]{@{}l@{}}Center-Line \\ Construction\end{tabular}    & \textit{\cmark} & \xmark          & \xmark          & \xmark          & \xmark          \\ \\
                                  & Object Detection                                                       & \textit{\cmark} & \textit{\cmark} & \xmark          & \textit{\cmark} & \textit{\cmark} \\ \\
                                  & Motion Planning                                                        & \textit{\cmark} & \textit{\cmark} & \textit{\cmark} & \textit{\cmark} & \textit{\cmark} \\ \\
                                  & Control Signal                                                         & \xmark          & \textit{\cmark} & \xmark          & \textit{\cmark} & \textit{\cmark} \\ \\
                                  & Lifelong Learning                                                      & \xmark          & \xmark          & \textit{\cmark} & \xmark          & \xmark          \\ \\
                                  & \begin{tabular}[c]{@{}l@{}}Collaborative \\ Driving\end{tabular}       & \xmark          & \xmark          & \textit{\cmark} & \xmark          & \xmark          \\ \\
                                  & Action Planning                                                        & \textit{\cmark} & \textit{\cmark} & \textit{\cmark} & \textit{\cmark} & \textit{\cmark} \\ \\
                                  & \begin{tabular}[c]{@{}l@{}}Adversarial \\ Attacks Defence\end{tabular} & \xmark          & \textit{\cmark} & \xmark          & \textit{\cmark} & \textit{\cmark} \\ \\
                                  & \begin{tabular}[c]{@{}l@{}}Reinforcement \\ Learning\end{tabular}      & \xmark          & \xmark          & \textit{\cmark} & \xmark          & \textit{\cmark} \\ \hline
\end{tabular}

 \label{my table}

\end{table}

\section{Opportunities and Challenges} \label{sec:op}

Applying ADVs in military domains provides numerous opportunities and streamlines operations such as terrain analysis, route optimization, medical evacuation, and border security without requiring human intervention. Integrating multimodal LLMs into the decision-making processes of ADVs helps meet military demands in areas such as transparency, adaptability, confidentiality, robustness, and the ability to handle vast amounts of data. Moreover, using multiple ADVs in military operations, communicating with each other and their surroundings via 6G communication, enhances perception by gathering data from various locations and enabling cooperative work. This reduces the workload on individual AV. The integration of 6G and edge computing, where edge devices, close to the data source, leverage collaborative approaches to share their resource and distribute the computation load of given tasks, further improves the robustness, scalability, and reliability due to keeping data on edge devices, of IoADV while reducing inference times in low resource environments, making it well-suited for real-time decision-making in the military domain. \cite{7}. Additionally, incorporating GenAI models optimizes the network, ensuring seamless communication. However, applying these technologies to the military domain faces some key challenges as listed below:

\subsection{Data Diversity, Reliability, and Compatibility}

The first challenge is ensuring redundancy and reliability in sensor systems to avoid catastrophic failures due to perceptions of the models and ADVs relying on these sensors since military operations can encounter dangerous situations that can harm the vehicle and its components. Another challenge is the compatibility of handling all these different data types in the same model. Different data types may exist in different dimensions and scales, causing the model to favor features with higher numerical magnitudes, and making it challenging to combine them into the single input data for the selected model.

\subsection{Training LLMs for Military Applications}
\textit{Data scarcity} is a significant issue, as military data is often sensitive and classified, limiting access to relevant datasets. This restricts the ability to develop models that accurately reflect the complexities of military language and scenarios. \textit{Classification issues} also arise due to the specialized terminology, acronyms, and context-dependent expressions prevalent in military communications. Misclassifications can lead to critical misunderstandings during operations, endangering personnel and mission success. To address these challenges, there is a growing need for synthetic data generation. By creating simulated datasets that mirror military contexts, researchers can enhance model training, particularly for rare or extreme scenarios.

However, military domains require complete verification for synthetic data to ensure that artificial datasets accurately represent real-world scenarios due to the reliability of the datasets for model training \cite{15}. One of the well-known verification methods is called statistical analysis, which ensures alignment in distributions, correlations, and variances. This comparison confirms that the synthetic data accurately reflects the characteristics of actual operational environments. If synthetic data is collected through the simulation, it is tested in simulated military scenarios using relevant operational parameters, such as battlefield topography, weather conditions, enemy tactics, and so on, to assess its realism. Moreover, incorporating military domain expertise is crucial for the effective validation of synthetic data and identifying potential inconsistencies in defense applications.

\subsection{Complete Real-Time Integration Processing}

In real-time applications like task-specific decision-making for ADVs, the model must generate actions within milliseconds. However, LLMs are computationally intensive and can experience high inference times. Additionally, handling diverse datasets requires significant pre-processing, which consumes time, and communication and synchronization among numerous connected ADVs and their surroundings may create network bottlenecks, further delaying the system. Military operations, particularly AV-driven tasks, are more delay-sensitive than typical applications due to their critical nature, requiring immediate responses. Therefore, ensuring minimal latency is crucial before deploying ADVs in this domain.

\subsection{Cybersecurity Concerns}
Security researchers have demonstrated that even state-of-the-art LLMs are far from secure, exhibiting vulnerabilities to critical attacks such as prompt injection and dataset poisoning. These weaknesses can lead to significant risks, particularly in sensitive applications like military operations.
The vulnerabilities of LLMs can pose grave dangers in military applications, where the stakes are exceptionally high. For example, if an adversary were to execute a successful data poisoning attack on an LLM used for intelligence analysis, the model could generate misleading assessments about enemy movements or capabilities. This misinformation could lead to flawed strategic decisions, resulting in operational failures or even loss of life.

In another scenario, consider a military AV equipped with an LLM for navigation and communication. If an attacker were to employ a prompt injection attack, the AV could be misled into executing commands that divert it into ambush zones or cause it to ignore critical threats. Such a breach could compromise mission integrity, endanger personnel, and reveal operational details to adversaries.
Connected vehicles in the network can suffer from various malicious attacks that are used for either information leakage or operation sabotage \cite{8682048}. For example, ADVs rely on an array of sensors such as cameras, LiDAR, RADAR, and so on to perceive their environment, and these sensors are vulnerable to various forms of manipulation and attacks, such as signal jamming and sensor spoofing. Moreover, fake updates or Interruptions during critical updates by malicious actors prevent ADVs from performing proper actions and harm the operation. Therefore, Secure communication methodologies should utilize military-grade encryption like AES-256 and quantum-resistant cryptographic techniques to prevent cyber threats and ensure mission-critical data integrity.

\subsection{LLM Inherited-Limitations}
One major risk of using LLMs in military applications is "hallucination," where the model produces outputs that are factually incorrect or entirely fabricated. This poses a serious threat in high-stakes battlefield situations, where decisions based on false information can have catastrophic results. In dynamic combat scenarios, if commanders depend on LLMs for real-time analysis, a hallucination about crucial details—such as enemy positions or friendly force status—could result in misguided decisions, leading to missed opportunities or unnecessary engagements. Mitigating hallucination risks in LLM models ensures the reliability and accuracy of their outputs and it can be provided by the several model adaptation techniques. Retrieval-augmented generation (RAG), which enhances models with external knowledge sources, ensures that AI outputs are grounded in factual data. Moreover, fine-tuning domain-specific data and adjusting pre-trained models with high-quality, domain-specific datasets aligns the LLM's responses with verified and contextually accurate knowledge. Automated Reasoning, which is a verification mechanism, and employing mathematical proofs to validate AI system outputs can enhance trustworthiness. \cite{friha.comst.2024}

\subsection{Cost of Deployment and Integration of 6G}

The deployment and integration of 6G and MEC for ADVs come with significant costs, which include infrastructure upgrades, such as installing 6G base stations, building edge data centers, and upgrading backhaul networks. ADVs also need advanced and rugged hardware including reinforced chassis, temperature-resistant electronics, and shock-proof sensors, to withstand extreme operational environments and compatible with advance communications, such as 6G-compatible modems, antennas, and edge computing modules. These systems along with high-end sensors and AI-driven software can increase vehicle costs. Operational costs such as network maintenance, energy consumption, and secure data transfer will further add to the overall investment. Additionally, military operations are performed not only in urban areas but also in primitive areas where there are fewer technological infrastructures in dangerous zones or environments where infrastructures failure due to hostile actions. %ADVs should also handle their tasks in these situations.

Though terahertz (THz) communications adaptation can provide reliability by providing uninterrupted connectivity, THz signals are highly susceptible to environmental factors such as atmospheric absorption, precipitation, and obstacles, which can significantly attenuate the signal strength, especially in harsh environments, including remote or combat zones. To mitigate these issues, adaptive communication strategies, such as real-time environmental sensing and dynamic beamforming, can be employed to intelligently adjust transmission parameters in response to changing conditions. On the other hand, in the case of infrastructure failure, fallback mechanisms such as multi-layer network redundancy, Satellite or high-altitude platform (HAP)-based backup communication, mobile ad-hoc networks (MANETs) and device-to-device (D2D) communication are practical solutions to maintain operational continuity, providing alternative sources and backup of the service for critical conditions \cite{8682048}.

\section{Conclusions} \label{sec:5}
This article explores integrating edge intelligence with 6G-enabled Internet of Automated Defense Vehicles (IoADV) in challenging environments, utilizing Large Language Models (LLMs) to enhance decision-making and communication. By processing multi-modal sensor data and applying advanced AI, these vehicles excel in complex conditions, offering improved efficiency, safety, and situational awareness. However, challenges like ensuring reliable IoADV communication and managing latency and computational demands persist. The proposed LLM-based framework addresses these issues, advancing autonomous military operations. Future research aims to minimize LLM-induced latency, enhancing real-time adaptability for ADVs.

 \vspace{-0.08in}

\section*{Acknowledgment }\label{sec:6}
This work is supported by the Natural Sciences and Engineering Research Council of Canada (NSERC) CREATE TRAVERSAL and NSERC DISCOVERY programs. The authors would like to thank Dr. Othmane Friha for sharing his valuable insights on the use of (M)LLMs with AVs.

\bibliographystyle{IEEEtran}
%\bibliography{biblio.bib}{}

% \printbibliography[title={Bibliography}] % Print the

\end{document}